# De-Risking Development in Sub-Saharan Africa: A Qualitative Study of Investment Dynamics in Angola

Carmen Berta C De Saituma Cagiza[*]

School of Social Science and Global Studies, University of Southern Mississippi,

*Abstract*

This study investigates how Development Finance Institutions (DFIs) contribute to de-risking development in Sub-Saharan Africa by shaping Foreign Direct Investment (FDI) flows and supporting sustainable economic transformation. Focusing on Angola as a representative case, the research draws on qualitative interviews with international development advisors, foreign affairs professionals, and senior public sector stakeholders. The study explores how DFIs mitigate investment risk, enhance project credibility, and promote diversification beyond extractive sectors. While DFIs are widely recognized as catalysts for private sector engagement, particularly in infrastructure, agriculture, and manufacturing, their effectiveness is often constrained by institutional weaknesses and misalignment with national development priorities. The findings suggest that DFIs play a crucial enabling role in fragile and resource-dependent settings, but their long-term impact depends on complementary domestic reforms, improved governance, and strategic coordination. This research contributes to the literature on development finance by offering grounded empirical insights from an underexamined Sub-Saharan context.

---
[*]carmen.cagiza@allusainvestment.holdings

# 1. Introduction

Sub-Saharan Africa (SSA) stands at a critical juncture in its pursuit of inclusive and sustainable development. Despite its vast natural wealth and expanding working-age population, structural barriers, ranging from weak institutions to limited infrastructure, continue to impede long-term economic transformation (Rodrik, 2007; UNECA, 2020). Angola exemplifies this paradox. As one of SSA's top oil exporters, the country has enjoyed periods of rapid GDP growth, yet those gains have not translated into broad-based improvements in human development or sectoral diversification (IMF, 2022).

In light of these challenges, Development Finance Institutions (DFIs) have emerged as increasingly important players in the region's development landscape. Mandated to foster private-sector engagement in high-risk markets, DFIs leverage public and multilateral capital to de-risk investment environments, improve bankability, and crowd in private capital where commercial finance is often scarce (Humphrey, 2017; OECD, 2022). Their role extends beyond financing: DFIs often bring policy expertise, technical assistance, and credibility that help bridge the gap between development objectives and investment feasibility.

At the same time, Foreign Direct Investment (FDI) remains a key mechanism for economic growth, valued for its potential to stimulate capital inflows, technology transfer, and employment generation. However, FDI's developmental impact in SSA has been uneven. Investments are frequently channeled into extractive industries, creating enclave economies with limited integration into local value chains (Asiedu, 2006; Borensztein et al., 1998). Scholars and practitioners alike have raised concerns that such patterns, coupled with governance deficits and regulatory uncertainty, restrict the transformative potential of FDI (Farole & Winkler, 2014; Moran, 2011).

This study investigates the perceived role of DFIs in influencing investment dynamics and promoting more inclusive, diversified development in Angola. Specifically, it examines how high-level stakeholders, ranging from international advisors to public-sector actors, interpret DFIs' contribution to shifting FDI flows beyond traditional sectors like oil and gas.

The central research question is:

To what extent are Development Finance Institutions perceived as effective in de-risking investment and enabling sustainable, diversified development in Angola?

This article draws on original fieldwork conducted for the author's doctoral dissertation, which explored the evolving role of DFIs in facilitating investment-led development in fragile, resource-dependent contexts. The interviews conducted in 2022 provided the empirical foundation for this article and have been re-analyzed and situated within the current policy discourse using recent secondary data and scholarly literature.

By addressing this question through a qualitative lens, the study offers grounded insights into how DFIs function within Angola's evolving political economy. It contributes to the growing discourse on development finance by highlighting the importance of institutional context, local agency, and international partnership in shaping investment outcomes in fragile and resource-dependent settings.

## 2. Literature Review

The question of how to de-risk development in fragile economies remains central to contemporary development discourse, particularly in SSA. Despite vast natural resource wealth and growing populations, many countries in the region continue to face difficulties in attracting and sustaining productive investment that supports long-term, inclusive growth. Within this context, FDI has long been promoted as a pathway to development, with the potential to bring capital, technology, and employment. Yet, the evidence from SSA remains mixed and often disappointing.

Classic economic theories have framed FDI as a driver of structural transformation (Borensztein et al., 1998), but empirical outcomes vary greatly based on institutional quality, sectoral focus, and governance effectiveness (Moran, 2011; Rodrik, 2007). Recent evidence further confirms that governance quality plays a decisive role in determining whether FDI contributes to inclusive growth across Sub-Saharan Africa, underscoring the importance of institutional coherence and accountability in fragile economies (Ofori & Asongu, 2021). In much of SSA, including Angola, FDI has flowed predominantly into capital-intensive extractive industries, with limited spillovers into broader economic development (Asiedu, 2006). This has given rise to enclave economies, isolated from local production networks and vulnerable to commodity price shocks.

Angola exemplifies these investment dynamics. While the country has attracted significant FDI since the end of its civil war in 2002, most inflows have centered around oil and gas. Despite contributing to macroeconomic growth and foreign reserves, these investments have not catalyzed structural diversification or inclusive development (IMF, 2019). Investment in agriculture, manufacturing, infrastructure, and human capital has remained weak, highlighting the need for mechanisms that can mitigate risk and direct capital toward strategic, high-impact sectors.

This has led to growing interest in the role of DFIs as specialized actors that can help de-risk investment in complex environments. DFIs, such as the African Development Bank, International Finance Corporation, Proparco, and the U.S. International Development Finance Corporation, seek to bridge market gaps by offering blended finance, risk guarantees, and technical assistance to attract private capital where it is most needed (Humphrey, 2017; OECD, 2022).

DFIs operate under dual mandates: to stimulate private sector development and to deliver measurable development impact. Their presence in post-conflict or institutionally weak economies is often framed as catalytic, making otherwise unviable projects feasible and encouraging investor confidence in markets perceived as too risky (Gabor, 2021). In Angola, DFIs have supported projects in infrastructure, agriculture, and entrepreneurship, offering a potential shift from aid dependency to investment-led growth.

However, the effectiveness of DFIs in fulfilling this catalytic role is contested. DFIs, some scholars argue, are increasingly operating like commercial lenders, prioritizing financial returns over developmental impact (Attridge & Engen, 2019). Others question whether their interventions are sufficiently aligned with national development priorities or whether they simply replicate donor preferences. These tensions raise important concerns about additionality, accountability, and the political economy of development finance.

In the Angolan case, these debates are especially relevant. While DFIs have played a visible role in recent investment efforts, the long-term developmental outcomes of their projects remain uncertain. Critical questions persist: Are DFI-supported investments contributing to diversification and institutional capacity building? Do their strategies reflect Angola's own development agenda, such as the Plano de Desenvolvimento Nacional? Or are they reinforcing existing asymmetries in the investment landscape?

Although existing research offers macroeconomic assessments of DFI performance, few studies investigate how DFIs are perceived by national stakeholders involved in development planning, investment policy, and international cooperation. The perspectives of these actors are vital to understanding the political and institutional dynamics that shape DFI engagement and its long-term impact.

This study addresses this gap by qualitatively exploring how DFIs are understood and evaluated by senior policymakers, foreign affairs professionals, and development advisors in Angola. By focusing on stakeholder perceptions, the research provides fresh insights into the investment dynamics underpinning Angola's development trajectory and the extent to which DFIs are contributing to a more resilient, diversified, and de-risked economic future.

## 2.1. Conceptual Framework

This study draws on a multi-disciplinary conceptual framework that integrates development finance theory, institutional economics, and political economy to examine how DFIs influence FDI flows and shape development trajectories in fragile, resource-dependent economies such as Angola. DFIs have increasingly been recognized as key actors in mitigating investment risk and addressing structural market failures by providing concessional finance, partial guarantees, and technical support, particularly in contexts that are underserved by commercial capital (Humphrey, 2017; OECD, 2022). Recent work highlights how DFIs are adapting these instruments to operate more effectively in fragile and conflict-affected states, where private capital is especially risk-averse (Gavas & Pleeck, 2021; Tyson & Mandelli, 2022). The framework builds on the premise that while FDI can be a powerful engine of economic transformation, facilitating capital accumulation, technological upgrading, and job creation, its developmental effectiveness is conditional on the quality of domestic institutions. Empirical work by Borensztein et al. (1998) shows that the positive impact of FDI depends on absorptive capacity, while Moran (2011) and Rodrik (2007) emphasize the centrality of governance, policy coherence, and regulatory integrity in determining whether FDI contributes to diversification or reinforces extractive dependencies.

Within this structure, DFIs are conceptualized as external catalytic agents. Their presence may serve not only to directly fund undercapitalized sectors but also to reduce perceived investment risk and signal project credibility to other financiers (te Velde & Warner, 2007). However, the extent to which DFIs achieve this catalytic function is mediated by institutional factors, namely, whether host governments possess the capacity and political will to integrate DFI-supported initiatives into coherent national development strategies (Farole & Winkler, 2014).

Figure 1 illustrates the conceptual model used in this study. DFIs influence the direction and composition of FDI, ideally supporting a transition away from extractive dominance toward more diversified and inclusive economic sectors. However, the link between DFI intervention and development outcomes is not automatic. Instead, it is contingent upon governance quality, policy alignment, and institutional readiness. The framework thereby underscores the importance of viewing DFIs not as standalone agents of change but as embedded participants in broader development coalitions.

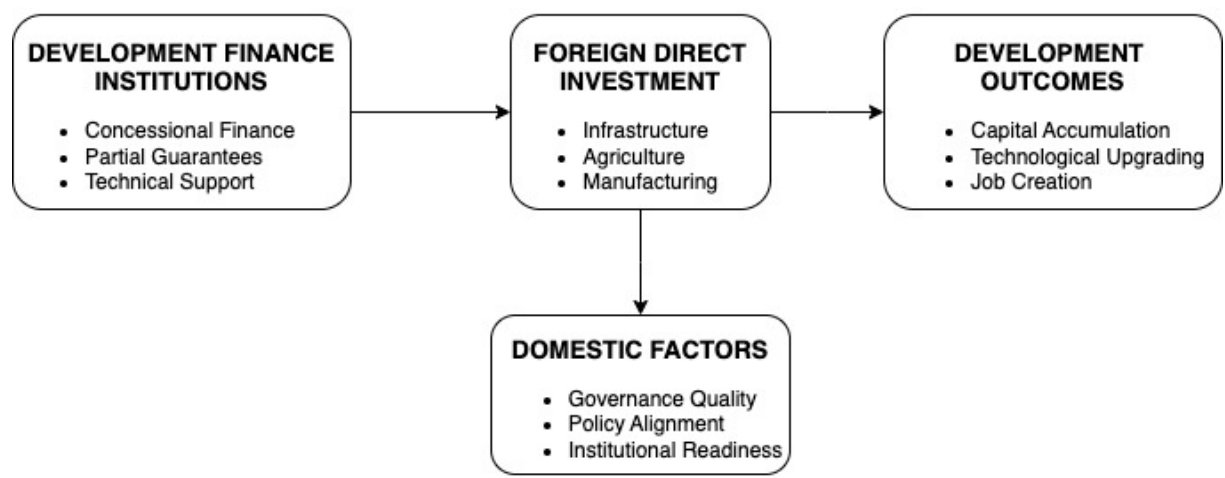

**Figure 1**. *Conceptual framework of the relationship between DFIs, FDI, governance conditions, and development outcomes in Angola*

## 3. Theoretical Framework

This study is anchored in two intersecting theoretical perspectives, Institutional Theory and Developmental State Theory, to critically examine how DFIs influence FDI and contribute to Angola's broader development objectives. These frameworks provide a structured lens through which stakeholder narratives are interpreted and contextualized within Angola's complex political economy.

Institutional Theory emphasizes the influence of both formal structures, such as legal frameworks, regulatory institutions, and governance norms, and informal practices in shaping economic decision-making and investment behavior (North, 1990; Scott, 2001). From this vantage point, DFIs do not act independently of context. Their success in catalyzing private investment and facilitating sectoral diversification is heavily contingent on the institutional environment of the host country. In fragile or resource-dependent states like Angola, institutional deficits, such as policy inconsistency, limited bureaucratic capacity, or weak rule of law, can undermine the catalytic ambitions of DFIs. Conversely, when DFIs enhance institutional credibility, improve transparency, or signal project viability, they may reduce perceived risk and encourage private sector participation.

Developmental State Theory complements this view by focusing on the state's proactive role in steering economic development through deliberate planning, strategic investment, and institutional coordination (Evans, 1995; Johnson, 1982). This theory is especially pertinent to Angola, where the state remains central to development planning, particularly in post-conflict reconstruction and economic diversification efforts. From this perspective, DFIs are not merely financiers but also development partners whose impact depends on the degree of alignment with national priorities. Their effectiveness is judged not solely by financial leverage, but by how well their interventions support or complement state-led initiatives aimed at long-term structural transformation.

Together, these theoretical lenses offer a comprehensive basis for interpreting the qualitative data. Institutional Theory explains how governance quality and institutional context mediate the effectiveness of DFIs, while Developmental State Theory provides insight into the strategic partnership potential between DFIs and the Angolan state. The application of these theories allows the study to explore not only the functional role of DFIs in mobilizing capital, but also their broader strategic influence on national development pathways. This integrated framework thus aligns directly with the study's aim: to assess how DFIs are perceived to shape FDI flows, enable diversification, and support inclusive development in Sub-Saharan Africa through the case of Angola.

## 4. Methodology

This study adopts a qualitative research design to explore the complex relationship between DFIs, FDI, and sustainable economic development in Angola. Qualitative methods are particularly well suited to uncover the social, institutional, and political dynamics embedded in development finance, which often elude capture through quantitative approaches (Creswell & Creswell, 2014; Patton, 2015). Angola's unique post-conflict context, heavy dependence on extractive industries, and evolving engagement with DFIs make it an appropriate case for an interpretive and exploratory inquiry.

### 4.1. Research Design and Rationale

This study employs a qualitative research design to explore the perceptions and experiences of stakeholders involved in development finance and investment planning in Angola. The decision to adopt a qualitative approach is grounded in the exploratory nature of the research objective and the complexity of the subject matter. Qualitative methods are particularly well suited for capturing how individuals interpret institutional dynamics, policy trade-offs, and strategic development interventions in contextually specific and politically sensitive environments (Creswell & Poth, 2018; Patton, 2015).

While the existing literature provides robust quantitative evidence on FDI flows and macroeconomic trends in Sub-Saharan Africa, far less attention has been devoted to understanding how DFIs are perceived by stakeholders embedded in national policy systems and public-sector institutions. This gap necessitates an interpretive approach capable of uncovering context-specific meanings and institutional narratives.

Semi-structured interviews were chosen for their flexibility and ability to generate detailed, comparative insights across participants. This method allowed respondents to elaborate on their views while enabling the researcher to probe emerging themes within a consistent thematic framework (Rubin & Rubin, 2012). The interview data captured not only participants' perceptions of DFIs as risk mitigators and strategic partners, but also their reflections on the tensions between donor-driven agendas and national development priorities.

By facilitating the collection of rich, descriptive data, this design made visible the institutional constraints, enabling conditions, and political realities often obscured in survey-based or econometric studies (Maxwell, 2013). Angola's post-conflict trajectory, dependence on extractive industries, and evolving engagement with development finance actors render it an especially relevant setting for qualitative inquiry into investment dynamics.

### 4.2. Sampling and Participants

A purposive sampling strategy was employed to identify participants with direct experience and expert knowledge of Angola's development finance and investment environment. A total of 14 senior-level participants with direct experience in development finance, investment policy, public sector strategy, multilateral cooperation, and international development were interviewed. Three interviews were conducted in Portuguese, acknowledging the linguistic and cultural context of key participants.

Participants were selected based on their strategic roles in advising or negotiating with DFIs, designing national investment strategies, or implementing sectoral reforms aligned with diversification and private sector engagement. While the sample size was intentionally limited, the depth of expertise yielded high-quality, information-rich insights. The aim was not statistical generalization, but rather an in-depth understanding of the perspectives and decision-making dynamics of those positioned to influence or assess DFI interventions.

The qualitative interviews analyzed in this study were originally conducted in 2022 as part of the author's doctoral dissertation on development finance and foreign investment in Angola. These interviews formed the basis for this article's core findings, offering timely insights into how stakeholders perceived DFIs as catalysts for FDI and broader structural transformation. The timing of the data collection coincided with key post-pandemic economic reforms and heightened DFI engagement, making the perspectives captured particularly relevant. While conducted in 2022, the findings remain analytically robust due to the continuity of Angola's macroeconomic context and institutional frameworks. The manuscript has also been updated with recent policy guidance and academic literature (2021–2024) to ensure contemporary alignment and strengthen its contribution to ongoing development finance debates.

### 4.3. Data Collection

Data were collected through semi-structured interviews conducted remotely using secure video and audio platforms. This format allowed flexibility in scheduling while ensuring confidentiality, an important consideration given the political sensitivity of the research context in Angola. Each interview lasted between 45 and 60 minutes.

The interview guide was informed by the literature and organized around four thematic domains: (1) patterns and distribution of FDI; (2) the catalytic and signaling role of DFIs; (3) institutional and governance challenges; and (4) alignment of DFI interventions with national development priorities. This semi-structured approach enabled open-ended dialogue while maintaining consistency across interviews (Rubin & Rubin, 2012).

### 4.4. Data Analysis

All interviews were audio-recorded with participants' consent and transcribed verbatim. A hybrid inductive-deductive thematic analysis was conducted following Braun and Clarke's (2006) framework. Initial open coding allowed for the emergence of unanticipated themes, while subsequent axial coding facilitated the clustering of data into higher-level categories. NVivo software was used to support the coding process and to conduct cross-case comparisons.

Thematic categories were refined iteratively and subsequently aligned with the study's conceptual framework to ensure analytical coherence. This approach enabled the triangulation of stakeholder narratives with existing academic literature, highlighting areas of convergence and divergence, and generating insights into the mechanisms through which DFIs shape investment behavior and development trajectories.

To further support the thematic analysis, a word cloud was generated using NVivo based on the coded transcripts (see Figure 2). This visual representation highlights the most frequently occurring terms across stakeholder interviews, including "investment," "risk," "DFIs," "governance," and "development." These terms closely align with the major themes identified in the coding process and reinforce the analytical saturation achieved in the study. The word cloud thus serves as an additional validation tool, illustrating the consistency between participant emphasis and the thematic structure of the findings.

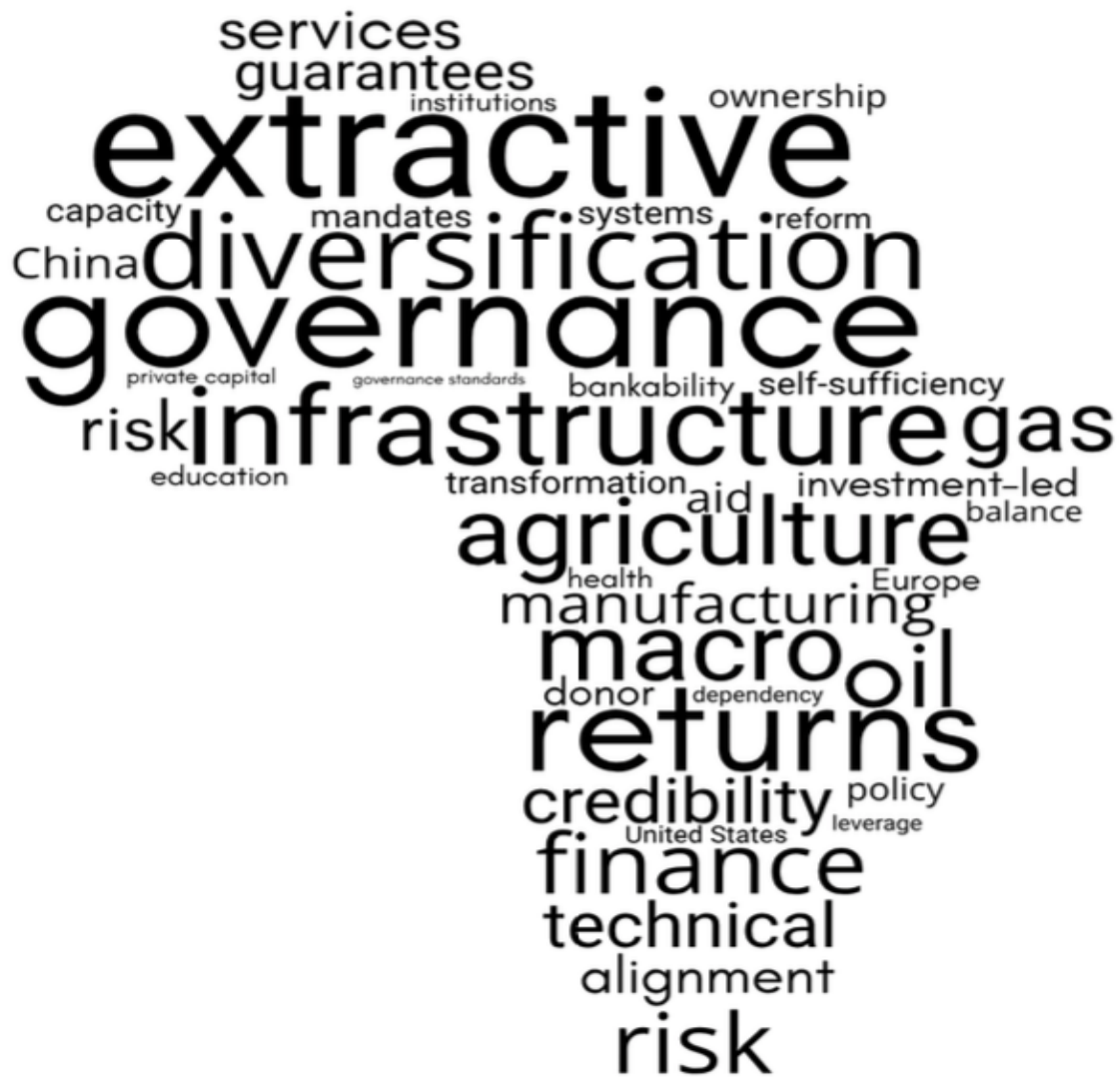

**Figure 2.** *Word Cloud of Key Terms Coded from Interview Transcripts*

This word cloud presents the most frequently occurring terms across 14 stakeholder interviews, reflecting dominant themes in perceptions of DFIs, investment dynamics, and institutional challenges in Angola. The visual reinforces the prominence of topics such as risk, governance, diversification, returns, and infrastructure, corroborating the thematic analysis and indicating saturation of core codes during data collection.

### 4.5. Ethical Considerations

The study followed internationally recognized ethical standards for research involving human subjects. All participants were informed of the research purpose, the voluntary nature of participation, and their rights, including the option to withdraw at any point. Verbal informed consent was obtained before each interview. No financial or material compensation was provided.

Three interviews were conducted in Portuguese and later translated into English for analysis and reporting. To preserve the accuracy and intent of the original responses, the translations were reviewed by native speakers fluent in both languages and familiar with development finance terminology. While minor interpretive nuances are inherent in any cross-language translation, steps were taken to minimize distortion or bias.

Given the seniority and political visibility of participants, extra care was taken to anonymize data. Identifying details, such as names, roles, or affiliations, were omitted from transcripts and quotations. Where appropriate, statements have been paraphrased to protect identities. These measures ensured participant confidentiality and encouraged openness during the interviews.

### 4.6. Study Limitations

As a qualitative study grounded in interviews with senior-level stakeholders, the findings offer contextually rich insights but are not statistically generalizable. The focus on senior stakeholders excludes grassroots perspectives, such as those of small entrepreneurs, local civil society actors, or direct DFI beneficiaries. Furthermore, given the political environment in Angola, some participants may have moderated their responses. Despite the methodological rigor and safeguards employed, these limitations should be acknowledged. Future studies could benefit from a mixed-methods approach or broader stakeholder engagement to validate and expand on these findings.

## 5. Results and Discussion

Thematic analysis of the qualitative interviews uncovered five interconnected themes that reveal how key stakeholders perceive the role of DFIs in influencing FDI and promoting sustainable development in Angola. These themes offer insight into the structural, institutional, and strategic dynamics that shape investment outcomes in a post-conflict, resource-dependent setting.

The five emergent themes are visually summarized in Figure 3, which illustrates how stakeholders perceive the role of DFIs in influencing FDI dynamics and promoting development in Angola. It synthesizes five interrelated themes derived from semi-structured interviews with senior policy, development, and investment stakeholders. The mind map highlights how DFIs are perceived to influence FDI outcomes across catalytic, geopolitical, sectoral, and institutional dimensions in Angola. It visually reflects the nuanced roles DFIs play in enabling or constraining investment-led development in fragile, resource-dependent contexts.

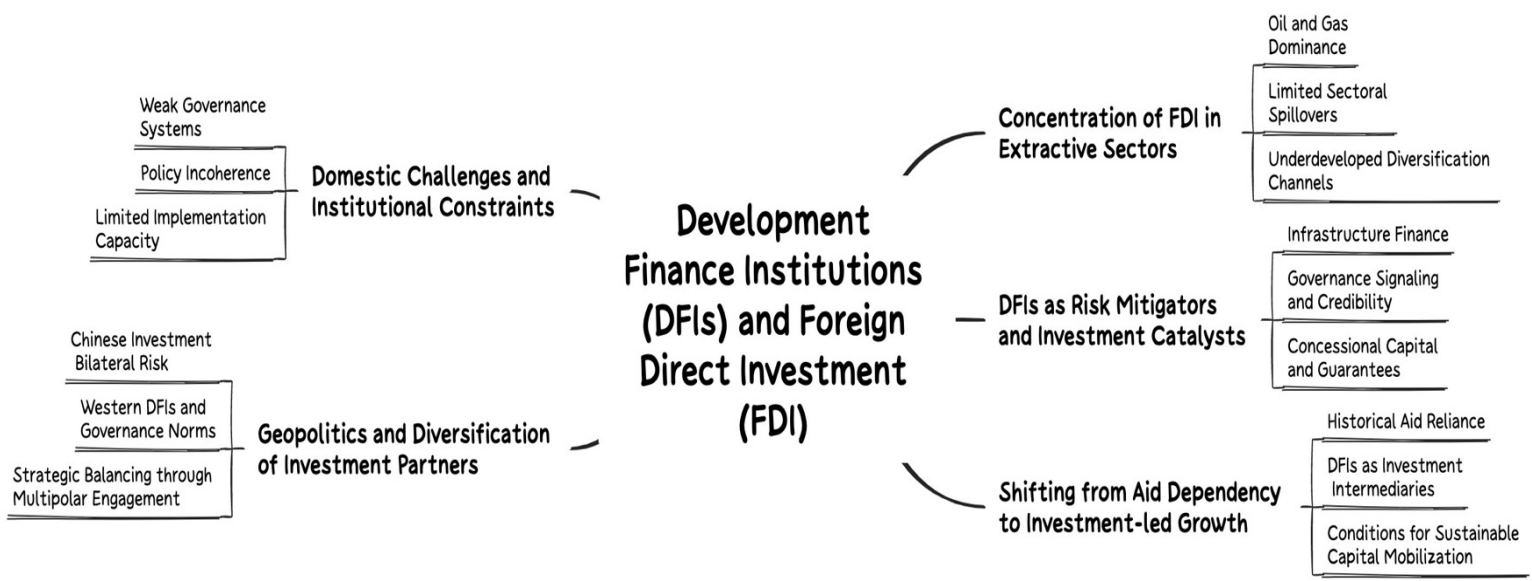

**Figure 3:** *Mind map of stakeholder perceptions of DFIs and FDI in Angola*

## 5.1. Concentration of FDI in Extractive Sectors

Participants consistently emphasized that Angola's FDI landscape remains heavily concentrated in oil and gas. While recognizing that these investments have driven macroeconomic growth, many noted that the broader development impacts have been modest. There was widespread concern that investment has not sufficiently diversified into sectors such as agriculture, manufacturing, or services.

As one foreign affairs professional observed:

"Foreign investors feel safe in oil and gas. The infrastructure is in place, and the returns are predictable. But this focus leaves other critical sectors struggling to attract interest."

This concern echoes existing literature that critiques the "enclave economy" model, where resource-rich countries experience capital inflows that generate limited domestic linkages or spillover effects (Asiedu, 2006). Although some interviewees acknowledged recent efforts to diversify FDI, most argued that institutional and policy reforms are still needed to redirect investment toward undercapitalized sectors that align with national development goals. This observation is echoed in post-pandemic data, where UNCTAD (2023) finds that while FDI in Sub-Saharan Africa has recovered, it remains heavily skewed toward extractives unless reinforced by blended finance and clear development mandates.

## 5.2. DFIs as Risk Mitigators and Investment Catalysts

The enabling function of DFIs emerged as a central theme. Participants viewed DFIs as essential actors that reduce risk perceptions and enhance project bankability in sectors typically overlooked by commercial financiers. Instruments such as concessional loans, risk-sharing guarantees, and technical assistance were seen as critical to mobilizing private investment in areas like infrastructure and agribusiness.

A senior public sector advisor commented:

"Without DFIs, many infrastructure or agro-industrial projects would stall. They bring not just capital but credibility."

This is consistent with recent sectoral studies, such as Dossou et al. (2023), which highlight that in renewable energy investments across Sub-Saharan Africa, governance quality remains a decisive factor in whether DFIs effectively mobilize private capital. It also reinforces arguments in the literature that DFIs can play a catalytic role by de-risking complex investments and signaling project quality (Humphrey, 2017). However, nearly 40% of respondents raised concerns that DFI interventions are not always well-aligned with Angola's development priorities. Some pointed to the influence of external mandates from parent institutions, which may result in project selection that prioritizes donor agendas over national needs.

### 5.3. Shifting from Aid Dependency to Investment-led Growth

A recurring theme was Angola's evolving development trajectory, from dependence on foreign aid to an investment-led approach focused on long-term economic transformation. While some participants acknowledged that development assistance played a stabilizing role in Angola's post-conflict recovery, many expressed skepticism about its long-term developmental value.

As one international development advisor stated:

"Aid helped us survive, but it won't help us transform. We need DFIs and FDI, but also stronger domestic governance to make that capital work."

This sentiment aligns with global debates around the limitations of aid-driven models and the need to mobilize private capital for development (UNCTAD, 2020). Interviewees viewed DFIs as important intermediaries that bridge public development objectives and private investment, though they emphasized that the broader impact of these investments depends heavily on the quality of domestic institutions and governance.

### 5.4. Geopolitics and Diversification of Investment Partners

The geopolitical dimensions of Angola's investment partnerships were also prominent in the interviews. Participants acknowledged the significance of Chinese investment in post-war reconstruction, particularly in infrastructure. However, many expressed concerns about the long-term implications, citing debt accumulation and limited knowledge transfer as drawbacks of these arrangements.

Several respondents advocated for closer ties with Western DFIs and development agencies, not only for financing but also for their emphasis on governance, transparency, and institutional reform.

"We've benefited from China's support, but we need balance. Stronger engagement with U.S. and European institutions gives us better leverage and higher standards," one advisor explained.

Emerging models led by African institutions, as documented by Convergence (2023), demonstrate how regionally tailored blended finance platforms can complement these partnerships by aligning more closely with domestic priorities and regulatory realities. This reflects a strategic interest in diversifying investment partners and avoiding overreliance on a single source of external capital. It also supports Bräutigam's (2009) argument that multipolar engagement strategies can enhance bargaining power and development outcomes for African states.

### 5.5. Domestic Challenges and Institutional Constraints

Despite the positive perceptions of DFIs, nearly all participants pointed to persistent domestic constraints as major barriers to translating investment into inclusive development. Weak institutional capacity, regulatory inconsistency, and governance challenges were repeatedly cited as limiting the effectiveness of both FDI and DFI-supported projects.

A public finance expert remarked:

"Even with DFI support, we won't succeed unless we strengthen our systems. Governance, education, and health matter just as much as financing."

This aligns with Rodrik's (2007) thesis that institutions are the primary determinants of sustained development. DFIs may offer critical support, but their impact depends on the host country's ability to absorb capital, enforce regulations, and implement reforms. Several participants recommended that DFIs more explicitly integrate institutional development goals into their project strategies to enhance long-term effectiveness.

**Table 1.** *Key Themes from Stakeholder Interviews on DFIs and FDI in Angola*

| Theme | Summary Insight | Illustrative Stakeholder View |
|---|---|---|
| Concentration of FDI in Extractive Sectors | FDI remains predominantly concentrated in oil and gas, limiting spillovers to other productive sectors. | "Investors feel safe in oil, but other sectors are neglected." |
| DFIs as Risk Mitigators | DFIs are widely seen as essential actors in reducing perceived investment risk and enhancing project credibility. | "Without DFIs, many infrastructure projects wouldn't happen." |
| Transition from Aid to Investment-led Growth | There is a clear stakeholder preference for shifting from aid dependency toward sustainable, investment-driven development. | "Aid helped us survive, but it won't help us transform." |
| Geopolitical Strategy | Stakeholders seek to diversify external partnerships, balancing China's presence with Western institutions. | "China is important, but U.S. and EU give us leverage." |
| Domestic Institutional Constraints | Governance weaknesses and bureaucratic inefficiencies continue to hinder the developmental impact of foreign and DFI-backed investments. | "Even with DFI support, we must fix our own systems." |

Table 1 provides a summary of these themes, including illustrative stakeholder quotes. Together, these themes reveal a nuanced picture of how DFIs are perceived within Angola's investment and development landscape. While widely viewed as necessary catalysts for mobilizing investment beyond the extractive sector, their success depends on alignment with national priorities, the quality of institutional frameworks, and the diversification of both sectors and partners. These insights offer valuable implications for both policymakers and development finance practitioners seeking to de-risk development and accelerate inclusive growth in fragile, resource-rich contexts.

## 6. Policy Implications

The findings of this study yield several policy-relevant insights for Angolan authorities, DFIs, and international development partners operating in fragile, resource-dependent environments. While DFIs have proven instrumental in catalyzing investment and de-risking capital-constrained sectors, their long-term developmental impact hinges on the policy, institutional, and strategic contexts within which they operate.

First, there is an urgent need for Angolan policymakers to prioritize institutional strengthening. Reforms that enhance legal clarity, regulatory coherence, and administrative transparency are critical for building investor confidence and enabling productive capital deployment. Efforts to reduce bureaucratic inefficiencies and combat corruption will be essential in improving the credibility of Angola's investment environment. As underscored by North (1990) and Acemoglu and Robinson (2012), the quality of institutions plays a central role in shaping economic trajectories. A more predictable and rules-based policy framework can also help align investment flows with national development goals and reduce the risk of capital flight or short-termism.

Second, DFIs should more deliberately align their interventions with country-owned development priorities. While the provision of concessional finance, guarantees, and advisory services remains valuable, stakeholders emphasized that these interventions must be tailored to local contexts. DFIs should actively engage with national planning agencies, integrate local knowledge into project selection, and ensure that investments complement, rather than substitute, domestic reform efforts. As noted by Humphrey (2017), and Massa & te Velde (2011), DFIs are most effective when their strategies reinforce local capacities and are embedded within national policy frameworks.

Third, Angola, and other resource-rich states in Sub-Saharan Africa, would benefit from a more diversified foreign investment strategy. Although Chinese investment has contributed significantly to infrastructure development, overdependence on any single bilateral partner poses risks related to debt sustainability, political alignment, and long-term value retention. A more balanced approach, one that engages a broader set of DFIs, multilateral organizations, and private investors, can mitigate geopolitical risk, foster competition, and support the diffusion of technology and expertise (Bräutigam, 2009).

Fourth, both DFIs and governments must invest in building local absorptive capacity. Human capital development, vocational education, and entrepreneurship support are foundational to ensuring that DFI-backed projects and FDI inflows translate into sustainable development outcomes. Without a capable workforce and an enabling ecosystem for local enterprise, the broader economic impact of external investment remains limited. Promoting linkages between foreign and domestic firms, especially within value chains in manufacturing, agribusiness, and renewable energy, will be essential to generate employment, enhance productivity, and strengthen economic resilience (UNCTAD, 2014).

In summary, de-risking development through DFIs requires more than financial engineering; it demands robust institutions, inclusive policy design, and strong partnerships between international actors and national stakeholders. For Angola, aligning external finance with local priorities and institutional capabilities will be pivotal in moving from extractive-led growth toward a more inclusive, diversified development model.

## 7. Conclusion

This study examined the strategic role of Development Finance Institutions (DFIs) in de-risking investment and supporting sustainable economic transformation in Angola, a resource-rich but institutionally constrained country in Sub-Saharan Africa. Through in-depth qualitative interviews with foreign affairs professionals, international development advisors, and senior public-sector stakeholders, the analysis illuminated how DFIs are perceived to influence investment behavior, policy orientation, and development outcomes.

The findings underscore that DFIs serve as crucial intermediaries in fragile markets, not only by mitigating financial risk but also by enhancing project credibility and signaling investor confidence. Their involvement has been particularly impactful in sectors often overlooked by private financiers, including agriculture, infrastructure, and value-added manufacturing. However, the catalytic potential of DFIs is not automatic. It is shaped by the broader institutional and policy environment in which they operate. Weak governance, inconsistent regulatory frameworks, and limited absorptive capacity remain key barriers to achieving deeper developmental impact.

Crucially, the study finds that DFIs are most effective when their efforts are closely aligned with national development strategies and supported by capable domestic institutions. In Angola's case, meaningful progress toward economic diversification and inclusive growth will require a coordinated effort that links external financing to systemic reforms, ranging from human capital development to institutional modernization.

Ultimately, DFIs are enablers, but not substitutes, for domestic development leadership. Their success in fostering resilient investment ecosystems depends on strategic alignment, institutional trust, and long-term policy coherence. For Angola and similar economies across Sub-Saharan Africa, the challenge lies not only in attracting capital, but in ensuring that it flows toward national priorities and delivers broad-based, sustainable benefits.

**Acknowledgments**

The author acknowledges the support of the University of Southern Mississippi, the World Bank Group, and the Permanent Mission of Angola to the United Nations. This article draws in part from research originally conducted for the author's PhD dissertation at the University of Southern Mississippi. The author further extends appreciation for the academic guidance and institutional support that made the original fieldwork and conceptual development possible.

Maxwell, J. A. (2013). *Qualitative research design: An interactive approach (3rd ed.).* SAGE Publications.

Miles, M. B., Huberman, A. M., & Saldaña, J. (2014*). Qualitative data analysis: A methods sourcebook (3rd ed.).* SAGE Publications.

Moran, T. H. (2011). *Foreign direct investment and development: Launching a second generation of policy research*. Peterson Institute for International Economics.

North, D. C. (1990). *Institutions, institutional change and economic performance.* Cambridge University Press.

Ofori, I. K., & Asongu, S. A. (2021). Foreign direct investment, governance and inclusive growth in Sub-Saharan Africa. *Foreign Trade Review, 56*(4), 445–467. https://doi.org/10.1177/00157325211002273

Organisation for Economic Co-operation and Development (OECD). (2022). *Scaling up blended finance in developing countries: Towards a policy roadmap.* OECD Publishing. https://www.oecd.org/dac/financing-sustainable-development/blended-finance-policy-roadmap.htm

Patton, M. Q. (2015). *Qualitative research & evaluation methods: Integrating theory and practice (4th ed.)*. SAGE Publications.

Rodrik, D. (2007). *One economics, many recipes: Globalization, institutions, and economic growth.* Princeton University Press.

Rubin, H. J., & Rubin, I. S. (2012). *Qualitative interviewing: The art of hearing data (3rd ed.)*. SAGE Publications.

Scott, W. R. (2001). *Institutions and organizations (2nd ed.).* Sage Publications.

te Velde, D. W., & Warner, M. (2007). *The use of subsidies to address market failures in the infrastructure sector*. Overseas Development Institute. https://doi.org/10.2139/ssrn.1025164
26

**Appendix A. Interview Protocol**

This protocol guided the semi-structured interviews conducted in March 2022. It ensured consistency across sessions while allowing flexibility for participants to elaborate on context-specific issues.

**Interview Mode:** Remote (via secure video/audio conferencing)

**Duration:** 45–60 minutes

**Informed Consent:** Verbal consent obtained prior to recording

**Confidentiality:** Anonymity guaranteed; no identifying information retained

**Recording:** Audio-recorded with permission; transcribed verbatim

**Appendix B. Interview Guide**

The interview guide was organized around four thematic domains related to the role of Development Finance Institutions (DFIs) and Foreign Direct Investment (FDI) in Angola.

**1. FDI Patterns and Trends**

In your view, what have been the dominant trends in Angola's FDI over the past two decades?

Which sectors attract the most foreign investment, and why?

What are the main barriers to diversifying FDI inflows?

**2. Role of DFIs**

How would you describe the role of Development Finance Institutions in Angola?

Have DFIs been effective in catalyzing investment in underserved sectors?

What specific contributions, financial, technical, or policy-related, have DFIs made?

**3. Aid, Investment, and Development Strategy**

How do you view the balance between aid and investment in Angola's development path?

To what extent have DFIs helped reduce reliance on foreign aid?

**4. Institutional and Political Considerations**

What challenges exist in aligning DFI investments with national priorities?

How do domestic governance issues affect the success of DFI-supported projects?

What institutional reforms are necessary to enhance the developmental impact of foreign investment?

**Appendix C. Thematic Coding Framework**

Thematic analysis was conducted using NVivo 12, employing both inductive and deductive coding strategies. The table below outlines the principal themes and associated subthemes that emerged from the qualitative data, along with illustrative coding labels.

| Main Theme | Subthemes | Illustrative Codes |
|---|---|---|
| *FDI Concentration* | Dominance of Oil and Gas; Weak Sectoral Linkages | "Extractive-centric", "Low domestic spillover", "Investor risk aversion" |
| *DFIs as Catalysts* | Risk Mitigation; Concessional Finance; Technical Support | "Guarantee instruments", "Signaling effect", "Feasibility gap financing" |
| *Investment vs. Aid* | Transitioning from Aid; Post-Conflict Financing Models | "Shift to investment", "Aid fatigue", "Donor influence decline" |
| *Geopolitical Dimensions* | China's Presence; Western DFIs; Multipolar Strategies | "Strategic hedging", "Sino-Angolan engagement", "Alignment with global norms" |
| *Institutional Barriers* | Governance Deficits; Capacity Constraints; Policy Incoherence | "Regulatory opacity", "Administrative bottlenecks", "Policy unpredictability" |

## Appendix D. Sample Interview Transcript (Excerpt)

**Excerpt from Interview with Public Sector Advisor (Anonymized)**

**Question:** How do DFIs contribute to Angola's diversification strategy?

**Response:**

*"They give credibility to projects. When a DFI is backing a project, it sends a message that due diligence has been done. That helps attract private investors who might otherwise be hesitant."*

**Question:** Are there challenges in ensuring these investments align with national needs?

**Response:**

*"Yes. Sometimes the goals are donor-driven. We need better coordination so DFIs support what Angola actually prioritizes."*

**Note:** Full transcripts are available upon request, subject to ethical review and confidentiality agreements.

## Appendix E. Interview Transcript (Portuguese)

Entrevista com um profissional do setor público em Angola, realizada em março de 2022. Esta transcrição foi traduzida para o inglês para fins de análise, mas aqui é apresentada no idioma original para referência.

**Entrevistador(a):** Obrigado por concordar em participar desta entrevista. Gostaria de começar perguntando como o senhor(a) vê o papel das Instituições Financeiras de Desenvolvimento (IFDs) em Angola atualmente.

**Entrevistado(a):** As IFDs são importantes porque assumem riscos que os investidores privados geralmente evitam. Sem elas, muitos projetos essenciais de infraestrutura e agricultura simplesmente não sairiam do papel.

**Entrevistador(a):** O senhor(a) acha que esses investimentos estão alinhados com as prioridades de desenvolvimento do país?

**Entrevistado(a):** Nem sempre. Às vezes, os projetos são mais voltados para os interesses dos doadores do que para as reais necessidades locais. É preciso haver mais diálogo entre as IFDs e os órgãos nacionais.

**Entrevistador(a):** E sobre o investimento estrangeiro direto? Quais setores o senhor(a) acha que mais se beneficiam atualmente?

**Entrevistado(a):** O setor petrolífero ainda domina. Temos infraestrutura e know-how nesse campo, mas é essencial diversificar. Precisamos atrair investimento estrangeiro para áreas como agricultura, indústria transformadora e energias renováveis.

**Entrevistador(a):** Há desafios específicos que dificultam essa diversificação?

**Entrevistado(a):** Claro. A instabilidade política, a burocracia e a falta de mão de obra qualificada são grandes obstáculos. Sem reformas institucionais, será difícil mudar o panorama.

**Entrevistador(a):** Muito obrigado por suas observações. Isso contribui significativamente para nossa pesquisa.

## Appendix F. Data Availability and Ethical Documentation

### F1. Data Availability Statement

In accordance with ethical research standards involving human participants, full interview transcripts and audio recordings are not publicly accessible due to confidentiality agreements and the sensitivity of the data. However, anonymized excerpts and thematic coding summaries may be made available upon reasonable request for academic verification or scholarly use. All requests must be submitted to the corresponding author and will be subject to institutional review and the preservation of participant anonymity.

**F2. Anonymized Consent Form Text (Verbal Script)**

The following script was used to obtain verbal informed consent prior to all interviews:

"Thank you for agreeing to participate in this interview. The purpose of this study is to understand how Development Finance Institutions influence foreign investment and economic development in Angola. Your participation is entirely voluntary. You may decline to answer any question and may withdraw from the study at any time. Your responses will be recorded for transcription purposes, but your identity will remain anonymous in any publication or presentation. Do you consent to participate under these conditions and for this conversation to be recorded?"

All participants provided verbal consent before the interviews commenced. No personally identifying information, such as names, job titles, or institutional affiliations—was included in the final transcripts or analytical documentation.